\documentclass[aps,prc,twocolumn,groupedaddress]{revtex4}

\usepackage{epsfig}
\usepackage{graphicx}
\usepackage{fancyhdr}
\usepackage{pslatex}   
\usepackage{color}
\usepackage{amsmath, amsthm, amssymb} 

\newcommand {\genbod} {{\sc genbod}~}
\newcommand {\pythia} {{\sc pythia}~}

\newcommand {\pp} {\mbox{$p+p$}~}
\newcommand {\AuAu} {\mbox{$Au+Au$}~}
\newcommand {\rootsNN} {\mbox{$\sqrt{s_{NN}}$}}

\begin{document}

\title {Conservation Laws and the Multiplicity Evolution of
Spectra at the Relativistic Heavy Ion Collider}

\author{Zbigniew Chaj\c{e}cki}
\email{chajecki@mps.ohio-state.edu}
\author{Mike Lisa}
\email{lisa@mps.ohio-state.edu}

\affiliation{Department of Physics, Ohio State University,
191 West Woodruff Ave, Columbus, OH 43210, USA}

\date{\today}

\begin{abstract}
  Transverse momentum distributions in ultra-relativistic heavy ion collisions
carry considerable information about the dynamics of the hot system produced.
  Direct comparison with the same spectra from $p+p$ collisions has proved
invaluable to identify novel features associated with the larger system, in
particular, the ``jet quenching'' at high momentum and apparently much stronger
collective flow dominating the spectral shape at low momentum.
  We point out possible hazards of ignoring conservation laws in the comparison of
high- and low-multiplicity final states.
  We argue that the effects of energy and momentum conservation actually dominate
many of the observed systematics, and that $p+p$ collisions may be much more
similar to heavy ion collisions than generally thought.
\end{abstract}


\maketitle

\section{INTRODUCTION}

\subsection{Heavy Ion Physics: Relying on Comparison}

The physics program at the Relativistic Heavy Ion Collider (RHIC) at Brookhaven National Laboratory is
remarkably rich, thanks to the machine's unique ability to collide nuclei from $^1{\rm H}$ to $^{197}{\rm Au}$,
in fully symmetric (e.g. \AuAu or \pp) to strongly asymmetric (e.g. $d+Au$) entrance channels, over an energy
range spanning more than an order of magnitude.  
The capability to collide polarized protons provides access to an entirely
new set of fundamental physics, not discussed further here.

Achieving the primary aim of
RHIC-- the creation and characterization of a color-deconfined state of matter and its transition back
to the confined (hadronic) state-- requires the full capabilities of RHIC.  In particular, comparisons of
particle distributions at high transverse momentum ($p_T$) from \AuAu and \pp collisions, probe the color-opaque
nature of the hot system formed in the collisions~\cite{Gyulassy:1990ye,Baier:2000mf,Adcox:2001jp}.
Comparison with reference $d+A$ collisions were
necessary to identify the role of initial-state effects in the spectra~\cite{Adams:2003im}.  Comparing anisotropic
collective motion from non-central collisions of different-mass initial states (e.g. \AuAu versus $Cu+Cu$)~\cite{Bhalerao:2005mm}
tests the validity of transport calculations crucial to claims of the creation of a ``perfect liquid'' at RHIC~\cite{Gyulassy:2004zy}.
Indeed, a main component of the future heavy ion program at RHIC involves a detailed energy scan, designed to
identify a predicted critical point in the Equation of State of QCD~\cite{Ritter:2006zz}.

The need for such systematic comparisons is not unique to RHIC, but has been a generic feature of all
heavy ion programs~\cite{Nagamiya:1988ge,Tannenbaum:2006ch}, from low-energy facilities like the NSCL (Michigan State),
to progressively higher-energy facilities at
SIS (GSI), the Bevatron/Bevalac (Berkeley Lab), AGS (Brookhaven), and SPS (CERN).  The nature of heavy ion physics
is such that little is learned through study of a single system.

\subsection{Bigger is better}

Despite the necessary attention to smaller colliding partners,
these comparisons are ultimately aimed at identifying novel aspects of collisions between the heaviest ions, in which
a highly excited {\it bulk system} might be created, with a sufficient number of degrees of freedom such that it may be described
thermodynamically-- e.g. in terms of pressure, temperature, energy density, and an Equation of State (EoS).
If the energy density of this system is sufficiently large (typically estimated at $\epsilon_{\rm crit}\sim 1$~GeV/fm$^{3}$~\cite{Gyulassy:2004zy})
and its spatial extent considerably larger than the color-confinement length $\sim$~1~fm, then a new state of matter--
the quark-gluon plasma (QGP)~\cite{Shuryak:1980tp}-- may be created.  Microscopically, such a state might be characterized by colored objects
(or something more complicated~\cite{Shuryak:2004tx}); macroscopically, it represents a region on the phase diagram
in which the EoS is distinctly different than for the hadronic phase~\cite{Karsch:2001cy}.

Ultra-relativistic collisions between the heaviest nuclei enjoy the additional advantage that finite-size effects are small,
due to high-multiplicity final states.  In a small system (e.g. final state of an $p+\bar{p}$ collision) a statistical analysis
of yields requires a canonical treatment, due to the conservation of discreet quantum numbers such as baryon number and strangeness~\cite{Becattini:1997rv}.
For larger systems, a grand canonical treatment is more common~\citep[e.g.][]{BraunMunzinger:1994xr}, with finite quantum-number effects absorbed into,
e.g. ``saturation factors''~\cite{Letessier:1993hi}.

Due to the large available energy $\sqrt{s}$ and final-state multiplicity, energy and momentum conservation effects on
kinematic observables (spectra, momentum correlations, elliptic flow) are generally small.
They are accounted for with correction factors~\cite{Danielewicz:1987in,Borghini:2000cm} or neglected altogether.

\subsection{Multiplicity evolution of single-particle spectra}

Detailed single-particle spectra (e.g. $d^2N/dp^2_T$) have been measured at RHIC, for a variety of particle types.
Often, the shape of the ``soft'' ($p_T \lesssim 2$~GeV/c) part of the spectrum is compared to hydrodynamic
calculations~\cite{Kolb:2003dz} or fitted to simple ``blast-wave''
parameterizations~\citep[e.g.][]{Retiere:2003kf} to extract the collective flow of the system.
The ``hard'' sector ($p_T \gtrsim 4$~GeV/c) is assumed to be dominated by the physics of the initial-state, high-$Q^2$
parton collisions and resulting jets.  The physics of the ``firm'' sector ($2\lesssim p_T \lesssim 4$~GeV/c) may be the
richest of all, reflecting the dynamics of the confinement process itself~\cite{Fries:2003vb}.

We would like to focus not so much on the single-particle spectra themselves, but on their multiplicity dependence.
Much has been inferred from this dependence.
In the soft sector, blast-wave fits to spectra from high-multiplicity final states (associated with central $A+A$ collisions)
indicate strong collective radial flow; the same fits to low-multiplicity final states-- including minimum bias \pp collisions--
appear to indicate much weaker flow~\cite{Adams:2003xp}.  This seems to confirm a common assumption
that \pp collisions are not sufficiently ``large'' to develop
bulk collective behaviour.

In the hard sector, one of the earliest and most exciting observations~\cite{Zajc:2001va,Adcox:2001jp} at RHIC was that 
the high-$p_T$ yield from high-multiplicity \AuAu collisions was suppressed, relative to appropriately scaled lower-multiplicity
$A+A$ or minimum bias \pp collisions.  This has been taken as evidence of energy loss of hard-scattered partons through a very
color-dense medium.  Meanwhile, the high-$p_T$ part of the spectrum from high-multiplicity \pp collisions appear {\it enhanced}
relative to low-multiplicity \pp collisions~\cite{Adams:2006xb}, again suggesting that a color-dense bulk system is not produced
in \pp collisions.

In this paper, we discuss the effects of energy and momentum conservation on the multiplicity evolution of
single-particle spectra at RHIC.  
Energy and momentum conservation-induced constraints
(EMCICs)~\footnote{In~\cite{Chajecki:2008vg}, 
we discussed energy and momentum conservation-induced {\it correlations} (EMCICs) in multiparticle distributions.
In the present manuscript, we discuss these very effects with the same formalism, but projected onto the single-particle distributions.
It is convenient and natural, then, to use the same acronym here, replacing ``correlation'' with ``constraint.''}
have been largely ignored in the analyses just mentioned, probably due to two reasons.
The first is the field's usual focus on the highest-multiplicity collisions, where such effects are assumed small;
it seems natural to compare analyses of such systems to ``identical'' ones of smaller systems, forgetting that
EMCIC effects play an ever-increasing role in the latter case.
Perhaps the more important reason is that EMCICs do not generate ``red flag'' structures on single-particle
spectra; this is in contrast to multi-particle correlation analyses, in which conservation law-induced correlations
may be manifestly obvious and have even been used to estimate the number of unmeasured neutral particles 
in high energy collisions~\cite{Foster:1973mt}.
Especially with the enhanced attention on precision and detail at the SPS and RHIC,
there has been increasing discussion of EMCIC effects in 2-particle~\cite{Borghini:2000cm,Chajecki:2008vg},
3-particle~\cite{Borghini:2006yk}, and $N$-particle~\cite{Borghini:2003ur}
observables.  Below, we show that EMCIC effects on single-particle spectra are also significant, and may even dominate their
multiplicity evolution.

\subsection{Organization of this paper}

Several authors~\citep[e.g.][]{Becattini:2007zn} have discussed finite-number effects in statistical models,
and many numerical simulations of subatomic collisions conserve energy and momentum automatically~\citep[e.g.][]{Drescher:2001hp,Werner:1995mx}.
However, as pointed out by Knoll~\cite{Knoll:1980ma}, our question-- to what extent do EMCICs {\it alone} explain the multiplicity evolution of
spectra?-- cannot be addressed from these simulations themselves, since dynamic and kinematic evolution are interwoven in these models.
Thus, in Section~\ref{sec:2}, we discuss a formalism based on Hagedorn's generalization of Fermi's Golden Rule, in which
dynamics and kinematics (phasespace) factorize.  This leads to a formula for finite-number effects on single-particle spectra, due
solely to kinematics, for a fixed dynamical (``parent'') distribution.

In Section~\ref{sec:test}, we test the extreme ansatz that {\it all} of the experimentally-measured multiplicity dependence of
single-particle spectra is due to EMCICs.  We will find surprising agreement with this ansatz in the soft sector ($p_T\lesssim 1$~GeV/c).
We will discuss that our formalism is on less firm footing, conceptually and mathematically, at much higher $p_T$.  Nevertheless, we explore
this regime as well.  We find that,
in the hard sector, the data from heavy ion collisions is clearly {\it not} dominated by EMCICs, though we point out that ignoring
EMCICs, especially for $p+p$ collisions, may be dangerous even at high $p_T$.  

In Sections~\ref{sec:discussion} and~\ref{sec:outlook}, we summarize and give an outlook for future studies.

\section{Effects of energy and momentum conservation on single-particle spectra}
\label{sec:2}
\subsection{A restricted phase space factor}
\label{sec:Fermi}

Changing the size (central versus peripheral ion collisions, $e+e$ collisions, etc) and energy of a collision system
will lead to different measured single-particle distributions, reflecting (1) possibly different physical processes driving
the system and (2) effects due to phase space restrictions.  To focus on changes caused by the latter, we consider
some Lorentz-invariant  ``parent'' distribution $\tilde{f}\left(p\right)\equiv 2E\frac{d^3N}{dp^3}$, driven by
some unspecified physical process, but {\it un}affected by energy and momentum conservation.  For simplicity,
we assume that all particles obey the same parent distribution.

In the absence of other correlations, the measured single-particle distribution is related to the parent according
to~\cite{Danielewicz:1987in,Borghini:2000cm,Borghini:2003ur,Chajecki:2008vg}
\begin{align}
\label{eq:EMCIC1}
&\tilde{f}_c\left(p_1\right)=\tilde{f}\left(p_1\right)\times \\
&   \frac{\int\left(\prod_{j=2}^N d^4p_j \delta\left(p^2_j-m^2_j\right)\tilde{f}\left(p_j\right)\right)\delta^4\left(\sum_{i=1}^N p_i-P\right)}
{\int\left(\prod_{j=1}^N d^4p_j \delta\left(p^2_j-m^2_j\right)\tilde{f}\left(p_j\right)\right)\delta^4\left(\sum_{i=1}^N p_i-P\right)}  \nonumber ,
\end{align}
where $N$ is the event multiplicity.
The integral in the numerator of Equation~\ref{eq:EMCIC1} represents the number of configurations in which the $N-1$ other particles counter-balance
$p_1$ so as to conserve the total energy-momentum $P$ of the event, and the denominator, integrating over all $N$ particles, is a normalization.

For $N\gtrsim 10$~\cite{Chajecki:2008vg}, one may use the central limit theorem to rewrite the factor in Equation~\ref{eq:EMCIC1}
as~\cite{Danielewicz:1987in,Borghini:2000cm,Borghini:2003ur,Chajecki:2008vg}
\begin{align}
\label{eq:EMCIC2}
&\tilde{f}_c\left(p_i\right) = \tilde{f}\left(p_i\right) \cdot \left(\frac{N}{N-1}\right)^2 \times \\
& \exp\left[-\frac{1}{2(N-1)}\left(
\frac{p^2_{i,x}}{\langle p_x^2 \rangle}+\frac{p^2_{i,y}}{\langle p_y^2 \rangle}+\frac{p^2_{i,z}}{\langle p_z^2 \rangle}
+\frac{\left(E_i-\langle E \rangle\right)^2}{\langle E^2 \rangle -\langle E \rangle^2}\right)\right] \nonumber ,
\end{align}
where
\begin{equation}
\label{eq:averages}
\langle p^n_\mu \rangle \equiv \int dp \tilde{f}(p) \cdot p_\mu^n 
\end{equation}
are average quantities
and we have set the average three-momentum $\langle p_{(\mu=1,2,3)} \rangle = P_{\mu=1,2,3}/N = 0$. 
We stress that what appears in Equation~\ref{eq:averages} is the parent distribution
$\tilde{f}$, not the measured one $\tilde{f}_c$.  Hence, for finite multiplicity $N$,
the averages $\langle p^n_\mu \rangle$ are not the measured ones, which we define as
\begin{equation}
\label{eq:MeasuredAverages}
\langle p^n_\mu \rangle_c \equiv \int dp \tilde{f}_c(p) \cdot p_\mu^n .
\end{equation}
See also the discussion in Appendix~\ref{sec:ApplicabilityAppendix}.

Since $p_T$ distributions are commonly reported, we would like to estimate EMCIC distortions to $p_T$ distributions, integrated over
azimuth and a finite rapidity bin centered at midrapidity.
As discussed in Appendix~\ref{sec:rapidityAppendix}, for the approximately boost-invariant distributions at RHIC~\cite{Adams:2003xp},
the measured and parent $p_T$ distributions are related by
\begin{align}
\label{eq:EMCfpt}
\tilde{f}_{c}\left(p_T\right) &= \tilde{f}\left(p_T\right) \cdot
    \left(\frac{N}{N-1}\right)^{2} \times  \\ \nonumber
&  \exp\left[
- \frac{1}{2\left(N-1\right)}
\left(\frac{2p_T^2}{\langle p_T^2\rangle}+\frac{\overline{p_z^2}}{\langle p_z^2\rangle} \right.\right.  \\ \nonumber
&\left.\left. + \frac{\overline{E^2}}{\langle E^2\rangle-\langle E\rangle^2}
- \frac{2\overline{E}\langle E\rangle}{\langle E^2\rangle-\langle E\rangle^2}
+ \frac{\langle E\rangle^2}{\langle E^2\rangle-\langle E\rangle^2}
\right)
\right] .
\end{align}
The notation $\overline{X}$ indicates the average of a $X$ over the rapidity interval used; see Appendix~\ref{sec:rapidityAppendix}
for details.  These
averages depend, of course, on $p_T$ and should not be confused with global averages $\langle X \rangle$
(Equation~\ref{eq:averages}) which characterize the parent distribution.

We would also like to emphasize the fact that since Equation~\ref{eq:EMCfpt} depends on 
the energy of the particle (not just momentum) it becomes clear that 
the EMCIC effects are larger on heavier particles at the same $p_T$. 
Thus we should expect that the proton spectra will be more suppressed 
than pion spectra.

In what follows, we find that 
ignoring the $\overline{p_z^2}/\langle p_z^2\rangle$ term 
does not affect our results, since the numerator is small for the narrow rapidity windows used here, and
the denominator is large.  In discussions below, we set this term to zero.

\subsection{Straw-man postulate of a universal parent distribution}
\label{sec:postulate}

Equations~\ref{eq:EMCIC1}-\ref{eq:EMCfpt} are reminiscent of Fermi's ``Golden Rule''~\cite{Fermi:1950jd,Hagedorn:1960xx},
in which the probability for making a particular observation is given by the product of the squared matrix element and a
quantity determined by available phase space.  The first term represented the underlying physical process.  In his original
statistical model~\cite{Fermi:1950jd}, Fermi originally assumed it to be a constant representing the volume in which emitted
particles were produced; this is equivalent to setting $\tilde{f}\left(p\right)$ constant in Equation~\ref{eq:EMCIC1}.
While surprisingly
successful in predicting cross sections and pion spectra~\citep[e.g.][]{Barkas:1957xx,Cerulus:1959xx}, the emission volume
required to describe the data was considered unrealistically large~\cite{Chamberlain:1959xx}.  Using the mean value theorem,
Hagedorn~\cite{Hagedorn:1960xx} generalized the theory so that the ``physics term'' is the interaction matrix element, suitably
averaged over all final states.

We wish to make no assumptions about the underlying physics (represented by $\tilde{f}$) driving
the observed spectrum $\tilde{f}_c$.  Rather, we wish to quantify the effect of changing the
multiplicity $N$, which appears in the phase space term.

In particular, in the following Section, we compare measured single-particle spectra for different event classes.

We \underline{postulate} that the parent distributions for, say classes $1$ and $2$, are the same ($\tilde{f}_1=\tilde{f}_2$).
By Equation~\ref{eq:averages}, this implies $\langle p_{\mu}\rangle_1 = \langle p_{\mu}\rangle_2 \equiv \langle p_{\mu}\rangle$.
In this case, the {\it only} reason that the observed spectra differ ($\tilde{f}_{c,1}\neq\tilde{f}_{c,2}$) is the difference
in ``multiplicity'' $N_1\neq N_2$; see Section~\ref{sec:postulateDetails} for a discussion of $N_1$.

To eliminate the (unknown) parent distribution itself, we will study the ratio of observed $p_T$ distributions, which,
by Equation~\ref{eq:EMCfpt} becomes
\begin{align}
\label{eq:ratio}
&\frac{\tilde{f}_{c,1}\left(p_T\right)}{\tilde{f}_{c,2}\left(p_T\right)} = 
    K\times\left(\frac{\left(N_2-1\right)N_1}{\left(N_1-1\right)N_2}\right)^{2} \times  \\ \nonumber
&  \exp\left[
\left(\frac{1}{2\left(N_2-1\right)}-\frac{1}{2\left(N_1-1\right)}\right)
\left(\frac{2p_T^2}{\langle p_T^2\rangle}+
 \right.\right.  \\ \nonumber
&\left.\left. + \frac{\overline{E^2}}{\langle E^2\rangle-\langle E\rangle^2}
- \frac{2\overline{E}\langle E\rangle}{\langle E^2\rangle-\langle E\rangle^2}
+ \frac{\langle E\rangle^2}{\langle E^2\rangle-\langle E\rangle^2}
\right)
\right] , 
\end{align}
where the constant $K$ is discussed at the end of Section~\ref{sec:postulateDetails}.  As mentioned at the end
of Section~\ref{sec:Fermi}, numerically unimportant terms in $p_z$ have been dropped.

Naturally, our postulate cannot be expected to be entirely correct; one may reasonably expect the mix of
physical processes in \pp collisions to differ from those in \AuAu collisions.
Nevertheless, it is interesting
to find the degree to which the change in single-particle spectra may be attributed {\it only} to
finite-multiplicity effects.  We will find that the postulate works surprisingly well in some
regions, and fails in others.  As we will discuss, both the success and failure raise interesting and surprising possibilities.

\subsection{Testing the postulate - how to treat the parameters}
\label{sec:postulateDetails}

By our postulate, the phase space factor affecting a $p_T$ distribution is driven by four quantities.
Three, $\langle p_T^2\rangle$, $\langle E^2\rangle$ and $\langle E\rangle$, characterize the parent distribution,
while $N$ is the number of particles in the final state.
In general,
increasing any one parameter decreases
the effect of phase space restrictions on the observed distributions.  
But what should we expect these values to be?
They should characterize the relevant system in which a limited quantity of energy and momentum is shared.  
They are not, however, directly measurable, and should only approximately scale with measured values, for at
least five reasons discussed here.

Firstly, the energy
and momentum is shared among measured and unmeasured (neutrals, neutrinos, etc.) particles alike
so that $N$ should roughly track the measured event multiplicity $N_{meas}$, but need not be identical to it.
Secondly, emission of resonances smears the connection between $N$ and $N_{meas}$; e.g. the emission of an omega meson which
later decays into ``secondary'' particles ($\omega\rightarrow\pi\pi\pi$) increments $N$ by unity, rather than three, as far
as other particles are concerned.
This latter consideration also affects the kinematic parameters $\langle p_T^2\rangle$, $\langle E^2\rangle$ and $\langle E\rangle$.
While energy and momentum are, of course, conserved in resonance decay, the aforementioned quantities, themselves, are not.  
Thus, one need not expect perfect correspondence between the appropriate kinematic parameters in Equation~\ref{eq:ratio}, and
the measured ones.

Thirdly, even restricting consideration to primary particles, it is unclear that all of them should be considered
in the relevant ensemble of particles sharing some energy and momentum.  In particular, for space-time extended
systems in high-energy collisions, the momentum extent of characteristic physics processes (e.g. string breaking) and
causality in an approximately boost-invariant scenario suggest that rapidity slices of roughly unit extent should be considered
separate subsystems~\cite{Borghini:2006yk}.  Of course, the total available energy
in any event is shared among {\it all} such subsystems; i.e. the midrapidity subsystem in one event will not have exactly
the same available energy as that in another event.  However, such fluctuations are to be expected in any case--
surely individual collisions will differ from one another to some extent.  Thus,
we repeat our interpretation of the four parameters $N$, $\langle p_T^2\rangle$, $\langle E^2\rangle$ and $\langle E\rangle$:
they characterize the scale, in energy and momentum, of the limited available phasespace to an $N$-particle subsystem.

Fourthly, Equations~\ref{eq:EMCIC1}-\ref{eq:ratio} are appropriate for fixed $N$, while
we will be comparing to measured spectra selected by measured charged-particle multiplicity.  Thus, $N$ would
inevitably fluctuate within an event class, even if we could ignore the above considerations. Naturally, high multiplicity events contribute to spectra more than low multiplicity events. Similarly, the average multiplicity in two-particle correlations is even more shifted to higher multiplicities.
 
Fifthly, as already mentioned in Section~\ref{sec:Fermi}, the kinematic parameters 
$\langle p_T^2\rangle$, $\langle E^2\rangle$ and $\langle E\rangle$ correspond to the parent distribution,
which will only correspond identically to the measured one in the limit of infinite multiplicity (i.e.
no EMCIC distortions).  See also the discussion in Appendix~\ref{sec:ApplicabilityAppendix}.

For all of these reasons, we will treat $N$, $\langle p_T^2\rangle$, $\langle E^2\rangle$ and $\langle E\rangle$
as free parameters when testing our postulate against data.  Our aim is not to actually measure these quantities
by fitting the data with Equation~\ref{eq:ratio}; this is good, since
our fits to the data only very roughly constrain our four parameters, as discussed in the next Section.
Rather, our much less ambitious goal is to see whether ``reasonable''
values of these parameters can explain the multiplicity evolution of the spectra.

To get a feeling for these values, we look at \pp collisions at \rootsNN=200~GeV, simulated by
the \pythia event generator (v6.319)~\cite{Sjostrand:2000wi}.  In the model, we can identify
primary particles, thus avoiding some of the issues discussed above.  However, the fact that
\pythia conserves momentum means that we access $\langle p_\mu^n \rangle_c$ as defined by Equation~\ref{eq:MeasuredAverages},
not the parameters of the parent distribution.  Nevertheless, a scale for our expectations may be set.
Table~\ref{tab:pythiaprim} summarizes the result for primary particles satisfying a varying 
cut on pseudorapidity where all particle decays where switched off in \pythia simulations.
The results from simulations when resonance decays were included in simulations are presented in Table~\ref{tab:pythiaall}.
These two tables gives us rough estimates of ranges of the total multiplicity and kinematic variables 
that one may expect. 
The bulk component of single-particle spectra is often estimated with Maxwell-Boltzmann
distributions, with inverse slope parameters in the range $T\sim 0.15 \div 0.35~GeV$.
Again, simply for rough guidance, we list Maxwell-Boltzmann expectations for our 
kinematic parameters in Table~\ref{tab:limits}, assuming pion-dominated system.

\begin{table}
\begin{tabular}{|c|c|c|c|c|c|c|c|}
\hline
$\eta_{max}$  &  $\quad\langle N \rangle\quad$ & $\quad\langle p_T^2  \rangle_c\quad$ & $\quad\langle p_z^2 \rangle_c\quad$ & $\quad\langle E^2 \rangle_c\quad$ & $\quad\langle E \rangle_c\quad$ \\
\hline
1.0 & 7.5 & 0.58 & 0.41 & 1.45 & 0.98  \\
\hline
2.0 & 13.4 & 0.59  & 2.81 & 3.89 & 1.57   \\
\hline
3.0 & 17.9 & 0.59  & 12.95 & 14.01 & 2.65 \\
\hline
4.0 & 21.5 & 0.59 & 82.45 & 83.55 & 5.13  \\
\hline
5.0 & 23.4 & 0.59  & 262.88  & 265.03 & 8.29  \\
\hline
$\infty$ & 23.6 & 0.59 & 275.23 & 276.4 & 8.48  \\
\hline
\end{tabular}
\caption{For a given selection on pseudorapidity $|\eta |<\eta_{max}$,
the number and kinematic variables for primary particles from a \pythia simulation of 
$p+p$ collisions at $\sqrt{s_{NN}}=200$~GeV are given.  Units are GeV/c or (GeV/c)$^2$, as appropriate.
100k events were used and all decays were switched off in simulations.
\label{tab:pythiaprim}}
\end{table}

\begin{table}
\begin{tabular}{|c|c|c|c|c|c|c|c|}
\hline
\hline
$\eta_{max}$  &  $\quad\langle N \rangle\quad$ & $\quad\langle p_T^2  \rangle_c\quad$ & $\quad\langle p_z^2 \rangle_c\quad$ & $\quad\langle E^2 \rangle_c\quad$ & $\quad\langle E \rangle_c\quad$ \\
\hline
1.0 & 16 & 0.20 & 0.11 & 0.40 & 0.44  \\
\hline
2.0 & 29 & 0.21 & 0.76 & 1.05 & 0.68  \\
\hline
3.0 & 39  & 0.21 & 3.5 & 3.8 & 1.2  \\
\hline
4.0 & 47  & 0.21 & 24 & 25 & 2.2  \\
\hline
5.0 & 51 & 0.22 & 88 & 89 & 3.7  \\
\hline
\end{tabular}
\caption{For a given selection on pseudorapidity $|\eta |<\eta_{max}$,
the number and kinematic variables for final state  particles (particle index KS=1 in \pythia) from a \pythia simulation of 
$p+p$ collisions at $\sqrt{s_{NN}}=200$~GeV are given. 100k events were generated and default 
\pythia parameters were used in simulations.  Units are GeV/c or (GeV/c)$^2$, as appropriate.
\label{tab:pythiaall}}
\end{table}

\begin{table}
\begin{tabular}{|c|p{1.8cm}|p{1.4cm}|p{3.4cm}|}
\hline
 &  non-rel. limit & ultra-rel. limit & if $T= 0.15 \div 0.35 GeV$ \\
\hline
$\quad\langle p_T^2  \rangle\quad$ &\quad$2mT$ &\quad$8T^2$ &\quad$0.045 \div 0.98~(GeV/c)^2$ \\
\hline
$\quad\langle E^2  \rangle\quad$ &\quad$\frac{15}{4}T^2 + m^2$ &\quad$12T^2$ &\quad$0.10 \div 1.50~GeV^2$ \\
\hline
$\quad\langle E  \rangle\quad$ &\quad$\frac{3}{2}T + m$ &\quad$3T$ &\quad$0.36 \div 1.00~GeV$ \\
\hline
\end{tabular}
\caption{The average kinematic variables obtained from the Maxwell-Boltzmann distribution $f(p)=\frac{dN}{dp^3} \sim e^{-E/T}$
using non-relativistic and ultra-relativistic limit. A pion gas is assumed.
\label{tab:limits}}
\end{table}

Finally, a word about normalization-- the quantity $K$ which appears in Equation~\ref{eq:ratio}.
Not only energy and momentum, but also discrete quantum numbers like strangeness and baryon number are conserved
event by event, affecting the overall yield of a given particle species.  For example, the related phenomenon of
``canonical suppression'' affects the ratio of yields for strange versus non-strange particles,
as multiplicity varies~\cite{Tounsi:2001ck,Fochler:2006et}.
Since we restrict our attention to energy and momentum conservation and the effect on kinematic quantities,
we are interested in the {\it shape} of the spectra ratio, as a function of
particle momentum, and include a factor $K$ in our Equation~\ref{eq:ratio}, which should be of order, but not
necessarily identical to, unity.  We do not discuss it further.

\begin{figure*}[t!]
{\centerline{\includegraphics[width=0.75\textwidth]{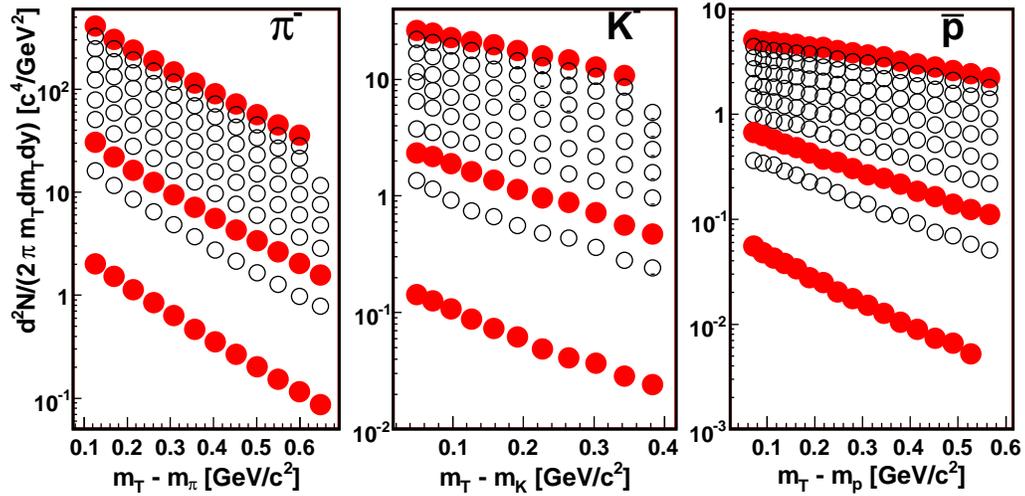}}}
\caption{(Color online)  Transverse mass distributions for pions (left), kaons (center) and antiprotons (right) measured by the STAR
Collaboration for \rootsNN=200~GeV collisions~\cite{Adams:2003xp}.  The lowest datapoints represent
minimum-bias \pp collisions, while the others come from \AuAu collisions of increasing
multiplicity.  Filled datapoints are for the top 5\% and 60-70\% highest-multiplicity \AuAu collisions, and for the \pp collisions.
\label{fig:STARPIDspectra}}
\end{figure*}

\begin{figure}[t]
{\centerline{\includegraphics[width=0.45\textwidth]{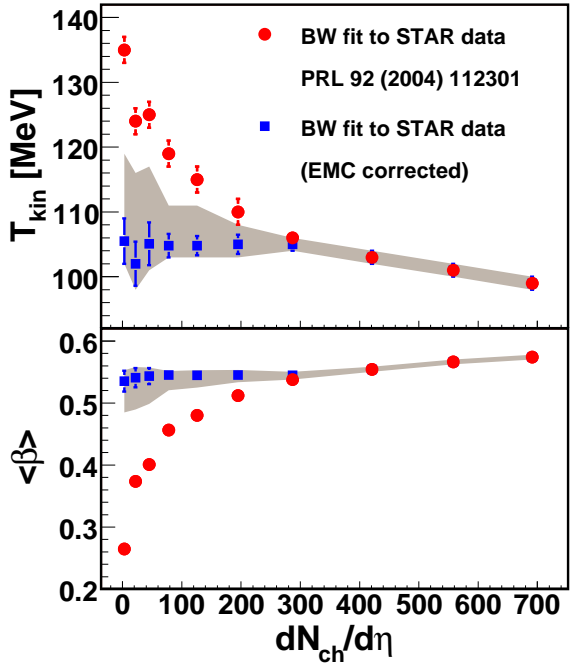}}}
\caption{(Color online) Circles show the temperature (top panel) and flow (bottom panel) parameters of
a Blast-wave model~\cite{Retiere:2003kf} fit to the STAR spectra of Figure~\ref{fig:STARPIDspectra},
as a function of the event multiplicity. Squares represent 
Blast-wave fit parameters to ``EMCIC corrected spectra,'' and shaded region
represents these results combined with systematic errors, as discussed in the text.
\label{fig:STARbeta}}
\end{figure}

\begin{figure}[t]
{\centerline{\includegraphics[width=0.45\textwidth]{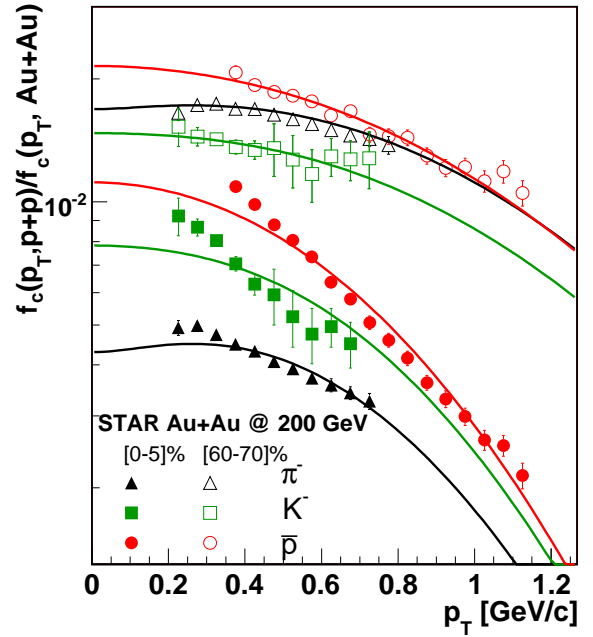}}}
\caption{(Color online) The ratio of the $p_T$ distribution from minimum-bias \pp collisions to the
distribution from 0-5\% (filled datapoints)
and 60-70\% (open datapoints) highest multiplicity \AuAu collisions; c.f. Figure~\ref{fig:STARPIDspectra}.
The ratio of the kaon spectra from \pp and 0-5\% \AuAu collisions (solid green squares) has been scaled by a factor 1.7 for clarity.
Curves represent a calculation of this ratio (ratio of EMCIC factors) using Equation~\ref{eq:ratio}.
\label{fig:SoftRatio}}
\end{figure}

\begin{figure*}[t]
{\centerline{\includegraphics[width=0.75\textwidth]{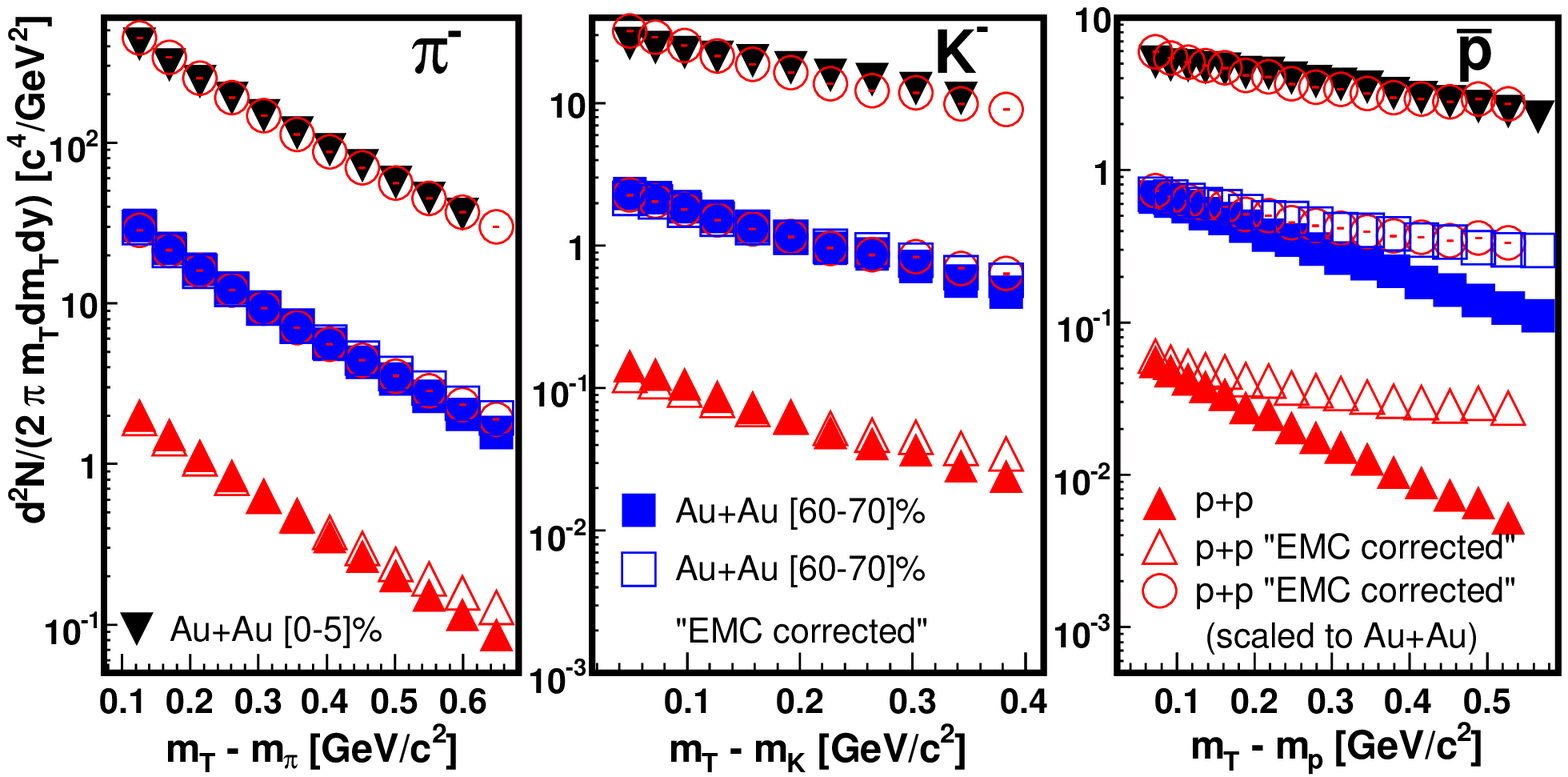}}}
\caption{(Color online) Transverse mass distributions of pions, kaons and antiprotons for minimum-bias \pp collisions
and 60-70\% and 0-5\% highest multiplicity \AuAu collisions at $\sqrt{s_{NN}}=200$~GeV.
Filled datapoints are the same as in Figure~\ref{fig:STARPIDspectra}.  Open triangles represent
the \pp spectra divided by the lower curves shown in Figure~\ref{fig:SoftRatio}.  Open circles
are the same spectra as the open triangles, except scaled up to compare to the spectra from the \AuAu
collisions.  Open squares represent the spectra from 60-70\% highest multiplicity \AuAu events,
divided by the ratio of upper and lower curves shown in Figure~\ref{fig:SoftRatio}.  See text for details.
\label{fig:EMCcorrectedSoftSpectra}}
\end{figure*}

\section{Test of the postulate - comparison to data}
\label{sec:test}

We now explore the degree to which the postulate proposed above describes
 the multiplicity evolution of measured $p_T$ spectra measured in $\sqrt{s_{NN}}=200$~GeV collisions at RHIC.
As is frequently done, we will separately discuss the ``soft'' ($p_T\lesssim 1$~GeV/c) and ``hard'' ($p_T\gtrsim 3$~GeV/c)
portions of the spectra.  This separation is not entirely arbitrary, as spectra in these two $p_T$ ranges are thought
to be dominated by quite different physics, and the multiplicity evolution in the two sectors is usually interpreted
in terms of distinct physics messages.  

In the soft sector, the spectral shapes are often consistent with hydrodynamic
calculations~\citep[e.g.][]{Kolb:2003dz,Huovinen:2006jp}, or fitted with blast-wave 
type models~\citep[e.g.][]{Schnedermann:1993ws,Retiere:2003kf}, and show evidence of
strong, explosive flow associated with a collective bulk medium.
This is especially clear in the mass dependence of the spectra; the $m_T$ (or $p_T$) spectrum of heavy particles like protons
are significantly flatter than that for pions, in the presence of strong flow.
The multiplicity evolution in this sector suggests that high-multiplicity collisions (say, central \AuAu collisions) show much more collective
flow than do low-multiplicity (say, \pp) collisions~\cite{Adams:2003xp}.  Such an interpretation initially sensible in a scenario in
which flow is built
up through multiple collisions among emitted particles; the concept of a collective bulk medium in a very low-multiplicity collision
is thus usually considered questionable.

Particle yields at high $p_T$, on the other hand, are generally discussed in the context of fragments from high-$Q^2$ parton
scatterings in the initial stage of the collision.  As the event multiplicity in \AuAu collisions is increased, a suppression
of high-$p_T$ yields is observed, relative to a properly normalized minimum-bias spectrum from \pp collisions.  This suppression
has been attributed to partonic energy loss in the bulk medium~\cite{Adams:2005dq,Adcox:2004mh,Back:2004je,Arsene:2004fa}.

The multiplicity evolution of the spectra in \pp collisions, however, shows quite the reverse.  Relative to the soft sector,
the high-$p_T$ yields increase as the multiplicity increases; one may also say that the $p_T$ spectra become less steep as
multiplicity increases~\cite{Adams:2006xb}.
This seems to reinforce the conclusion discussed above in relation to the soft sector, that \pp collisions do not build up
a bulk system capable of quenching jets.

Here, we reconsider these conclusions based on 
 the multiplicity evolution of the spectra, in light of the phase space restrictions discussed above.

\subsection{Soft sector: identified particles in \AuAu versus \pp}
\label{sec:olga}

Figure~\ref{fig:STARPIDspectra} shows $m_T$ distributions for minimum-bias \pp collisions and multiplicity-selected
\AuAu collisions, all at $\sqrt{s_{NN}}=200$~GeV, reported by the STAR Collaboration at RHIC~\cite{Adams:2003xp}.
For the highest-multiplicity \AuAu collisions (top-most filled datapoints), the spectrum for heavier emitted particles
is less steep than the essentially exponential pion spectrum.  Circles in Figure~\ref{fig:STARbeta} show the result of
fits with a blast-wave model~\cite{Retiere:2003kf}.  They indicate a kinetic freezeout
temperature of about 100~MeV and average collective flow velocity about 0.6c for the most central collisions.
For lower multiplicity collisions, the freeze-out temperature appears to grow to $\sim 130$~MeV and the flow velocity 
decreases to $\sim 0.25\rm{c}$.
The STAR collaboration, using a slightly different implementation of a blast-wave model, reported essentially identical values~\cite{Adams:2003xp}.

Ratios of spectra from minimum-bias \pp collisions to those from \AuAu collisions are plotted in Figure~\ref{fig:SoftRatio}.
For the filled points, the denominator is the most central \AuAu collisions, while the open points represent the ratio
when the denominator is from peripheral (60-70\% centrality) \AuAu collisions.
Pions, kaons, and protons are distinguished by different symbol shapes.

The curves show the function given in Equation~\ref{eq:ratio}, for the kinematic scales given in Table~\ref{tab:soft}.
Clear from the Table is that all curves in Figure~\ref{fig:SoftRatio} are generated with the
same kinematic variables $\langle p_T^2\rangle$, $\langle E^2\rangle$ and $\langle E\rangle$; only
the relevant multiplicity changes.

\begin{table}[t]
\begin{tabular}{|l|c|c|c|c|}
\hline
Event selection & N & $\langle p_T^2\rangle$ [(GeV/c)$^2$] & $\langle E^2\rangle$ [GeV$^2$] & $\langle E\rangle$ [GeV] \\
\hline
\hline
\pp min-bias         & 10.3  & 0.12 & 0.43 & 0.61 \\ 
\AuAu 70-80\%         & 15.2  &  "   &  "   &  "   \\
\AuAu 60-70\%         & 18.3  &  "   &  "   &  "   \\
\AuAu 50-60\%         & 27.3  &  "   &  "   &  "   \\
\AuAu 40-50\%         & 38.7  &  "   &  "   &  "   \\
\AuAu 30-40\%         & 67.6  &  "   &  "   &  "   \\
\AuAu 20-30\%         & 219 &  "   &  "   &  "   \\
\AuAu 10-20\%         & $> 300$  &  "   &  "   &  "   \\
\AuAu 5-10\%          & $> 300$  &  "   &  "   &  "   \\
\AuAu 0-5\%           & $> 300$  &  "   &  "   &  "   \\
\hline
\end{tabular}
\caption{Multiplicity and parent-distribution kinematic parameters which give a reasonable 
description of the spectrum ratios for identified particles in the soft sector.  See text
for details.  Note that the multiplicity changes with event class; the parent distribution
is assumed identical.
}
\label{tab:soft}
\end{table}

We do not quote uncertainties on the kinematic or multiplicity parameters,
as the fitting space is complex, with large correlations between them.
Furthermore, it is clear that the calculated curves do not perfectly reproduce the measured ratios.
However, it is also clear that ``reasonable'' values of multiplicity and energy-momentum scales go
a long way towards explaining the
multiplicity evolution of the spectra, even keeping physics (``parent distribution'') fixed.  Our postulate of 
Section~\ref{sec:postulate} seems to contain a good deal of truth.

Another way to view the same results is useful.
While the curves shown in Figure~\ref{fig:SoftRatio} only approximately describe the data shown there, 
one may approximately ``correct'' the measured $m_T$ distributions, to account for EMCICs.  This is shown in 
Figure~\ref{fig:EMCcorrectedSoftSpectra}, where the measured min-bias \pp and central and mid-peripheral \AuAu
spectra have been copied from the full points of Figure~\ref{fig:STARPIDspectra} and are shown by full points.  
The open red triangles represent the min-bias \pp spectra, divided by Equation~\ref{eq:ratio}, with the parameters
from Table~\ref{tab:soft}.
This ``EMCIC-corrected''
spectrum is then scaled up to show comparison to the spectra from central \AuAu (open red circles); 
the level of (dis)agreement is identical to that between the lower datapoints and curves in Figure~\ref{fig:SoftRatio}.

Spectra from the mid-central \AuAu collisions have been likewise ``corrected.''
The open squares in Figure~\ref{fig:EMCcorrectedSoftSpectra}
may be compared to the open circles;
again the level of (dis)agreement is equivalent to that between the upper datapoints and curves in Figure~\ref{fig:SoftRatio}.

\begin{figure*}[t]
{\centerline{\includegraphics[width=0.75\textwidth]{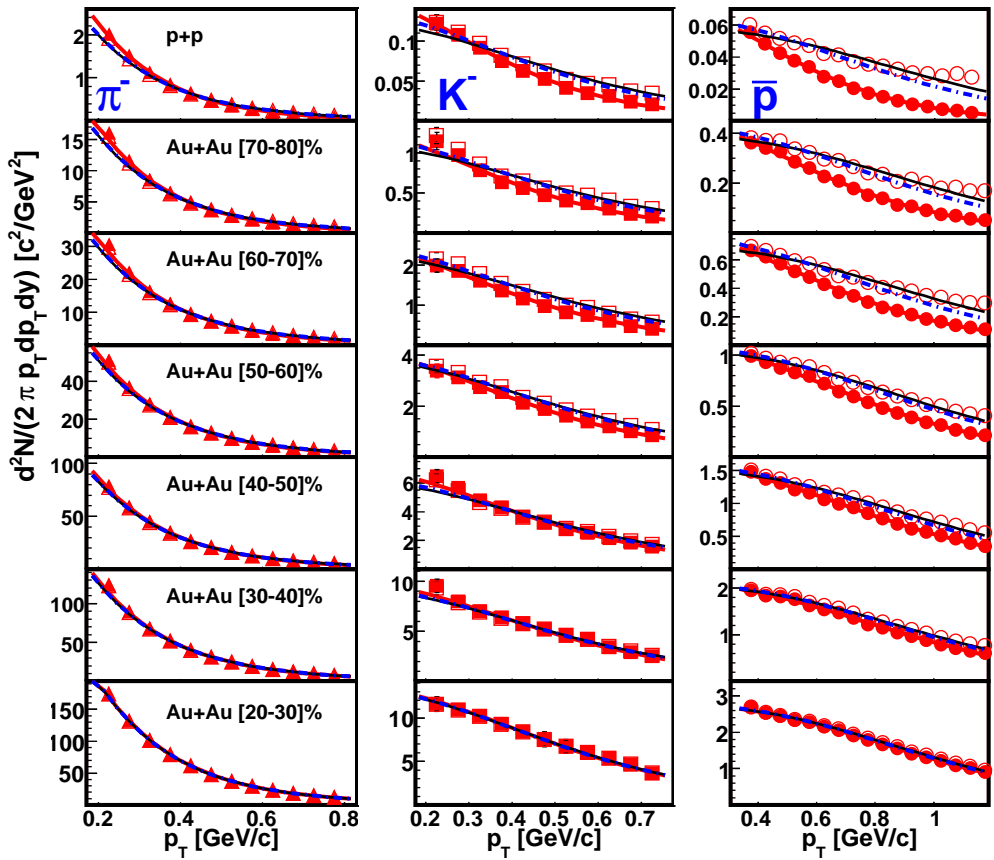}}}
\caption{(Color online) $dN/dp^2_T$ spectra for pions (left), kaons (center), and protons (right) are plotted on a linear scale, as a
function of event multiplicity.  Top panels show spectra for minimum-bias \pp collisions, and the spectra for the six
lowest multiplicity selections of \AuAu collisions are shown in the lower panels.  Filled symbols are the measured
data, while open symbols are the ``EMCIC corrected'' distributions, discussed in the text.  (For pions, these distributions
overlap almost completely.)  Blast-wave fits are indicated by the curves.  For the ``EMCIC corrected'' spectra, two fits
are performed, to estimate systematic errors.  The solid line represents a fit to all datapoints,
while the fit indicated by the dashed line ignores proton yields above $p_T=0.8$~GeV/c.
\label{fig:BWfitsLinear}}
\end{figure*}

Spectra themselves contain more information than two-parameter fits to spectra.  However, much has been made of blast-wave fits
to measured $p_T$ spectra, which suggest a much larger flow in central \AuAu collisions, relative to \pp collisions.  Thus, it may
be instructive to see how EMCICs affect these parameters.  In Figure~\ref{fig:BWfitsLinear}, the $p_T$ distributions for \pp
collisions and the six lowest multiplicity selections on \AuAu collisions are shown.  Blast wave fits to the measured spectra,
resulting in the parameters shown by red triangles in Figure~\ref{fig:STARbeta} are shown as curves.  On the linear scale of the
Figure, some deviations between the fit and data, particularly at the lowest $p_T$ for the light particle, is seen.  This has
been observed previously in Blast-wave fits, and may be due to resonances~\cite{Broniowski:2001we,Retiere:2003kf}.
Nevertheless, the fits to measured data are reasonable overall, and for simplicity, we do not exclude these bins.

Also shown in Figure~\ref{fig:BWfitsLinear} are the ``EMCIC corrected'' spectra, as discussed above.  As already seen in
Figure~\ref{fig:EMCcorrectedSoftSpectra}, these differ from the measured spectra mostly for low multiplicity
collisions and for the heavier emitted particles.  Blast-wave fits to these spectra are also shown.  Especially
for the very lowest multiplicity collisions, these fits are less satisfactory than those to the measured spectra;
the ``parent distributions'' extracted via our approximate EMCIC correction procedure follow the Blast-wave
shape only approximately.  Much of the deviation is at $p_T\sim 0.9$~GeV/c for protons from the lowest multiplicity
collisions (upper-right panels).  
This is the region around which the approximations used in deriving the EMCIC correction should start to break down,
as discussed in Appendix~\ref{sec:ApplicabilityAppendix}.
So, two fits are performed: one including all datapoints
shown (blue squares in Figure~\ref{fig:STARbeta}), and the other excluding proton spectra points with $p_T > 0.8$~GeV/c.  The resulting range of Blast-wave parameters
is indicated by the shaded region in Figure~\ref{fig:STARbeta}.  There, statistical errors on the fit parameters have
been multiplied by $\sqrt{\chi^2/{\rm d.o.f.}}$ (ranging from $\sim 2$ for spectra from \pp collisions to $\sim 1$
for those from mid-peripheral and central \AuAu collisions) and added to both ends of the range.  Thus, the shaded
region should represent a conservative estimate of blast-wave temperature and flow strengths to the parent distributions.

In summary,
to the extent that the curves in Figure~\ref{fig:SoftRatio} describe the ratios shown there-- which they
do in sign, magnitude and mass dependence, but only approximately in shape-- the data is consistent with
a common parent distribution for spectra from all collisions.
The residual deviation seen in Figure~\ref{fig:SoftRatio} is observed again in different forms in Figures~\ref{fig:EMCcorrectedSoftSpectra}
and~\ref{fig:STARbeta}.
The upshot is that EMCICs may dominate the multiplicity evolution of the spectra in the soft sector at RHIC.
Extracting physics messages from the changing spectra, while ignoring kinematic effects of the same
order as the observed changes themselves, seems unjustified.

In particular, STAR~\cite{Adams:2003xp} and others~\cite{Retiere:2003kf} have fitted the spectra with
Blast-wave distributions, which ignore EMCIC effects. Based on these fits, they concluded that 
the difference in spectral shapes between high- and low-multiplicity collisions was due to much
lower flow in the latter; c.f. Figure~\ref{fig:STARbeta}. Recently, Tang et al.~\cite{Tang:2008ud}
arrived to the same conclusion, using a modified Blast-wave fit based on Tsallis statistics.
This requires introduction of an extra parameter, $q$, intended to account for system fluctuation
effects~\cite{Wilk:2008ue}.
However, contrary to the claims in the Tang paper, the Tsallis distribution - with or without $q$ -
does {\it not} account for energy and momentum conservation~\cite{WilkPrivate}; EMCIC effects 
would need to be added on the top of the Tsallis statistics~\cite{WilkPrivate}. Therefore,
conclusions about flow in low-multiplicity collisions based on these fits are suspect.

An independent measurement of flow would help clarify this issue. Two-particle femtoscopy (``HBT'')
is a sensitive probe of collective motion~\cite{Lisa:2005dd} and has been measured in \pp collisions
at RHIC~\cite{Chajecki:2005zw}. Any scenario should be able to describe simultaneously both the spectral
shapes and the $m_T$ dependence of the femtoscopic scales. A study of this topic is underway.

\subsection{Soft sector: unidentified particles in multiplicity-selected \pp collisions}

\begin{figure}[t]
{\centerline{\includegraphics[width=0.45\textwidth]{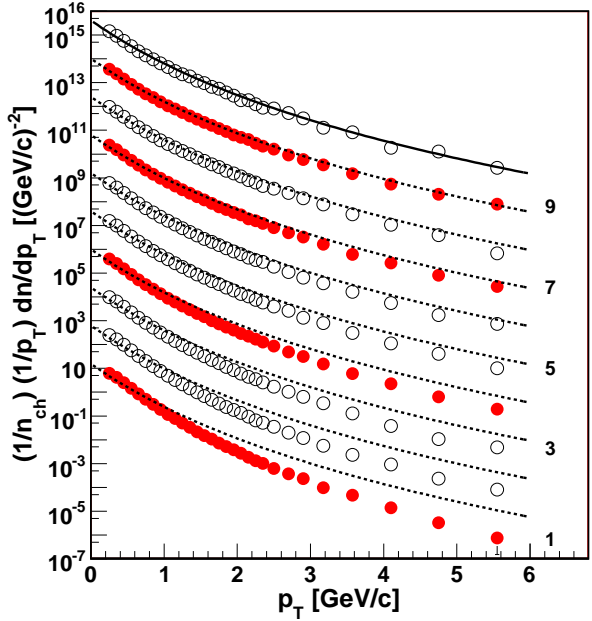}}}
\caption{(Color online) Transverse momentum spectra of unidentified negative hadrons from \pp collisions
at $\sqrt{s_{NN}}=200$~GeV by the STAR Collaboration~\cite{Adams:2006xb}.  The lowest (highest)
dataset corresponds to the lowest (highest) multiplicity collisions.  The solid line is intended
only to guide the eye and show the shape of the spectrum for the highest multiplicity selection.
It is rescaled and redrawn as dashed lines below, to emphasize the multiplicity evolution of the spectrum shape.
\label{fig:STARnonidSpectra}}
\end{figure}

While minimum-bias \pp collisions are the natural ``reference'' when studying \AuAu collisions, the
STAR experiment has also measured $p_T$ spectra from multiplicity-selected \pp collisions~\cite{Adams:2006xb}.
These are reproduced in Figure~\ref{fig:STARnonidSpectra}, in which the lowest-multiplicity collisions are
shown on the bottom and the highest at the top.  Numerical labels to the right of the spectra are included
just for ease of reference here.

The solid curve is a power-law fit to the highest-multiplicity spectrum (\#10), just for reference.  This curve
is scaled and replotted as dashed lines, to make clear the multiplicity evolution of the spectra.
Concentrating on the soft sector for the moment, we perform the same exercise as above, to
see to what extent this multiplicity evolution can be attributed to EMCICs.

\begin{figure}[t]
{\centerline{\includegraphics[width=0.45\textwidth]{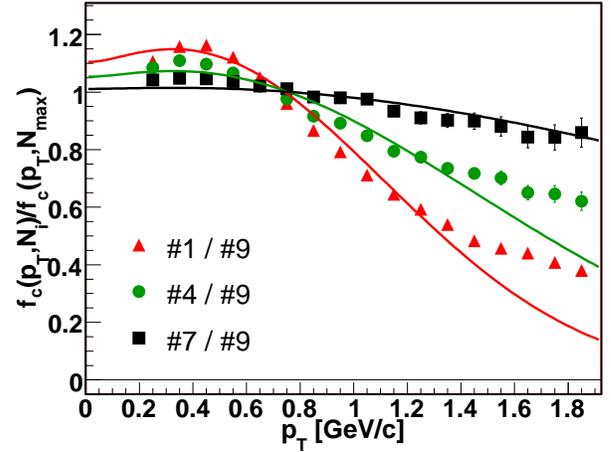}}}
\caption{(Color online) Ratio of the $p_T$ spectra shown by full points in Figure~\ref{fig:STARnonidSpectra}.
Spectra for the lowest-multiplicity (red triangles), fifth-lowest (green triangles) and seventh-lowest (squares)
multiplicity collisions are divided by the spectrum for the second-highest multiplicity collisions.
Curves represent a calculation of this ratio (ratio of EMCIC factors) using Equation~\ref{eq:ratio}; see text for details.
\label{fig:nonidSpectraFitsZoom}}
\end{figure}

In Figure~\ref{fig:nonidSpectraFitsZoom} are shown three ratios of spectra, in which the second-highest-multiplicity
spectrum (\#9) is used as the denominator, to avoid statistical fluctuations associated with the highest multiplicity
spectrum.
Also shown are curves, using Equation~\ref{eq:ratio} with the energy-momentum scales given in Table~\ref{tab:pp}.

\begin{table}[b]
\begin{tabular}{|l|c|c|c|c|}
\hline
Multiplicity cut & N & $\langle p_T^2\rangle$ [(GeV/c)$^2$] & $\langle E^2\rangle$ [GeV$^2$] & $\langle E\rangle$ [GeV] \\
\hline
\hline
\# 1         & 6.7  & 0.31 & 0.90 & 0.84 \\ 
\# 4         & 11.1 &  "   &  "   &  "   \\
\# 7         & 24.2 &  "   &  "   &  "   \\
\# 9         & 35.1 &  "   &  "   &  "   \\
\hline
\end{tabular}
\caption{
Multiplicity and parent-distribution kinematic parameters which give a reasonable 
description of the spectrum ratios for unidentified particles in the soft sector from multiplicity-selected \pp collisions.
See text for details.  Note that the multiplicity changes with event class; the parent distribution
is assumed identical.}
\label{tab:pp}
\end{table}

The spectra reported by STAR are for unidentified negative hadrons.
In calculating these curves, we assumed that all particles were pions.  This matters, since the
energy terms in Equation~\ref{eq:ratio} require the particle mass.  We expect the energy-momentum
scales listed in Table~\ref{tab:pp} to be affected by this simplistic assumption.  Particle-identified
spectra from multiplicity-selected \pp collisions would be required, to do better.
Given this, and the only semi-quantitative agreement between the calculations and measured ratios
shown in Figure~\ref{fig:nonidSpectraFitsZoom}, we conclude only that the EMCIC contribution to the
multiplicity evolution of low-$p_T$ spectra in \pp collisions is at least of the same order as the
observed effect itself.

\subsection{Segue: From the soft to the hard sector}

Figure~\ref{fig:SoftRatio} shows the central result of this paper: namely, that the multiplicity evolution of
the mass and $p_T$ dependence of single particle spectra in the soft sector may be
understood almost entirely in terms of phase-space restriction with decreasing event
multiplicity.

Plotted in that figure is the ratio of spectra from low-multiplicity events over spectra from
high-multiplicity events.  Experimental studies sometimes show this ratio's inverse, 
often called $R_{AA}$~\cite{Adcox:2001jp}.  While of course the same information is shown
in both representations, we choose that of Figure~\ref{fig:SoftRatio} for two reasons.
The first is to emphasize the effects of EMCICs, the topic of this paper;
these are, generically, to suppress the
particle yield at high energy and momentum, particularly for low-N final states.  
(In multiparticle distributions, they also
generate measurable correlations~\cite{Chajecki:2008vg}.)

The second reason is to stress that we have been discussing spectra in the soft sector, whereas
the ratio $R_{AA}$ is generally studied at high $p_T$.  At large $p_T$, we expect that a purely
EMCIC-based explanation of the multiplicity evolution of the spectra might break down, for
two reasons.
  Firstly, even if particles of all momenta shared phase-space statistically, our 
approximation of Equation~\ref{eq:EMCIC2} is expected to break down for energies
much above the average energy, as discussed in Appendix~\ref{sec:ApplicabilityAppendix}.
  Secondly, it is believed that the high-$p_T$ yield has a large pre-equilibrium
component; thus, high-$p_T$ particles might participate less in the statistical sharing of
phase-space, as discussed in Section~\ref{sec:postulateDetails}.  

As we discuss in the next Section, EMCICs surely do {\it not} dominate the multiplicity
evolution of the hard sector in heavy ion collisions.  
For interpreting high-$p_T$ spectra from multiplicity-selected \pp collisions, accounting 
for EMCICs may or may not be important.
In order to make the connection to Figure~\ref{fig:SoftRatio},
we will plot spectra from low-multiplicity collisions over those from high-multiplicity, as
well as the inverse, to make the connection to $R_{AA}$.

\subsection{Spectra in the hard sector}
\label{sec:hardSpectra}

The generic effect of EMCICs is to suppress particle yields at energy-momentum far
from the average value.  The effect is stronger for lower multiplicity $N$.  It is clear, then, that EMCICs
cannot account for the multiplicity evolution of the spectra at high $p_T$ in \AuAu collisions,
since high-multiplicity collisions are observed to have {\it more} suppression at high $p_T$ than do low-multiplicity collisions~\cite{Adcox:2001jp}.
Thus, we conclude that our postulate fails for \AuAu collisions at high $p_T$; the ``parent distribution'' describing
the underlying physics in this region does, indeed, change with multiplicity.

But in \pp collisions, the multiplicity evolution in the hard sector is opposite to that in \AuAu collisions.
In particular, in \pp collisions, the yield at high $p_T$ (relative to lower $p_T$) is {\it in}creased as multiplicity
increases, as is clear from Figure~\ref{fig:STARnonidSpectra}; similar results have been observed in $\overline{p}+p$ collisions
at the Tevatron~\cite{Alexopoulos:1988na}, ISR~\cite{Breakstone:1986xs}, and Sp$\overline{\rm p}$S~\cite{Arnison:1982ed}.
A ``hardening'' of the spectrum with increasing multiplicity goes in the same
direction as would EMCIC effects.  To what extent can EMCICs account for the multiplicity evolution of
spectra from \pp collisions, in the hard sector?

\begin{figure}[t]
{\centerline{\includegraphics[width=0.45\textwidth]{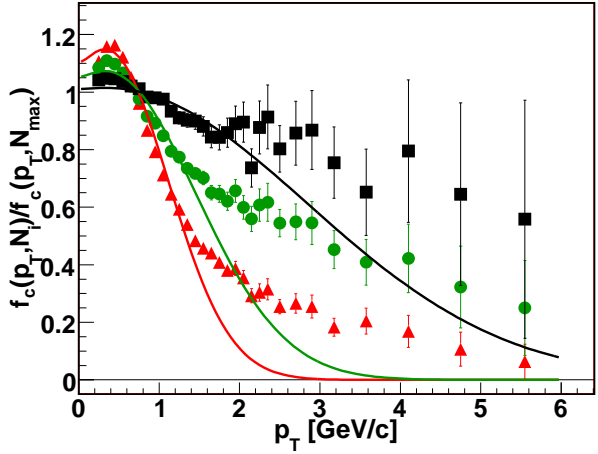}}}
\caption{(Color online) The same data and curves as shown in Figure~\ref{fig:nonidSpectraFitsZoom}, but plotted over the
entire measured $p_T$ range.
\label{fig:nonidSpectraFit}}
\end{figure}

Some insight on this question may be gained from Figure~\ref{fig:nonidSpectraFit}, in which the data and curves shown in
Figure~\ref{fig:nonidSpectraFitsZoom} are plotted out to $p_T=6$~GeV/c.  Clearly, the calculated suppression function
(Equation~\ref{eq:ratio}) fails dramatically at high $p_T$.

We recall that
Equations~\ref{eq:EMCIC2} and~\ref{eq:ratio} are based on
the central limit theorem (CLT), which naturally leads to Gaussian distributions.  As discussed in
Appendix~\ref{sec:ApplicabilityAppendix}, one expects
the breakdown of the CLT approximation in the far tails of the distribution-- e.g. when $p_T^2 \gg \langle p_T^2\rangle$.
Thus, any inferences we make about EMCIC effects in the hard sector remain qualitative.  Nevertheless,
the level of disagreement between the calculations and measurements leads us to conclude 
that EMCICs do {\it not} fully explain the multiplicity evolution of $p_T$ spectra in \pp collisions
in the hard sector.

\begin{figure}[t]
{\centerline{\includegraphics[width=0.45\textwidth]{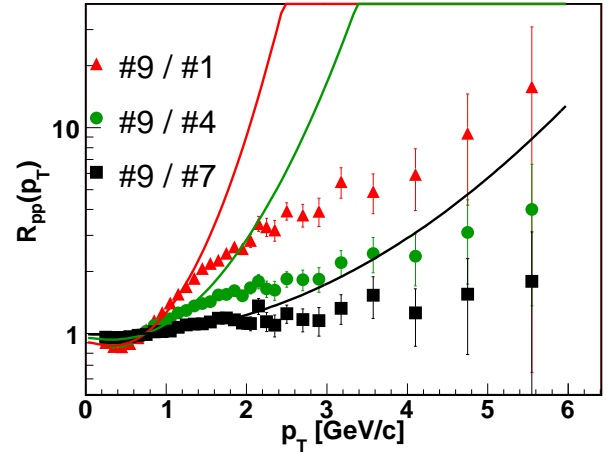}}}
\caption{(Color online) ``$R_{pp}$,'' the analogue of ``$R_{CP}$'' used in heavy ion collisions.  The spectrum from the highest-multiplicity
\pp collisions are divided by spectra from lower-multiplicity collisions (see filled datapoints in Figure~\ref{fig:STARnonidSpectra}).
The data and curves are simply the inverse of those shown in Figure~\ref{fig:nonidSpectraFit}.
\label{fig:Rpp}}
\end{figure}

However, this, in itself, raises a fascinating possibility.  Figure~\ref{fig:nonidSpectraFit}
shows that, relative to high-multiplicity \pp collisions,
the suppression of high-$p_T$ yields from low-multiplicity collisions is not as strong as one expects 
from our simple postulate.
Said another way, the high-$p_T$ ``enhancement'' in high-multiplicity collisions
may not be as large as one expects from phasespace considerations alone.  This is emphasized in Figure~\ref{fig:Rpp},
in which is plotted ``$R_{pp}$'', the ratio of the spectrum from high-multiplicity to lower-multiplicity collisions;
$R_{pp}$ is the analog of $R_{CP}$ from heavy ion collisions~\cite{Adcox:2001jp}.

The motivation for studying quantities like $R_{AA}$ and $R_{CP}$ (and now $R_{pp}$) is to identify important
differences between one class of collisions and another.  Presumably, one is interested in physics
effects (jet quenching, etc.), above and beyond ``trivial'' energy and momentum conservation.  Thus, it makes
sense to attempt to ``correct'' for EMCICs by dividing them out as we did in Section~\ref{sec:olga}, keeping in mind the
caveats just discussed.

\begin{figure}[t]
{\centerline{\includegraphics[width=0.45\textwidth]{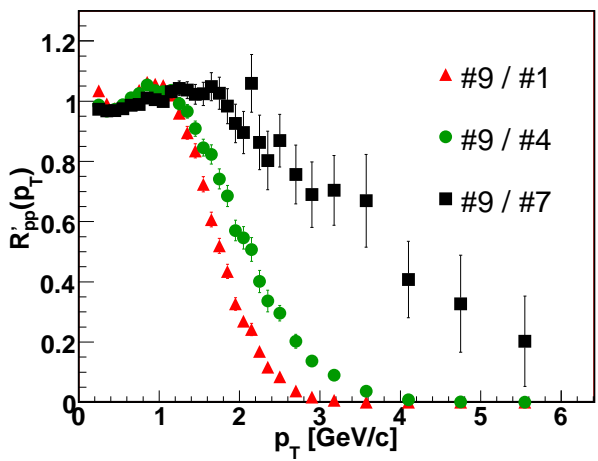}}}
\caption{(Color online) $R_{pp}$ (c.f. Figure~\ref{fig:Rpp}) divided by the EMCIC contribution to $R_{pp}$, as calculated by Equation~\ref{eq:RppPrime}.
\label{fig:EMCcorrectedRpp}}
\end{figure}

The result of this exercise is shown in Figure~\ref{fig:EMCcorrectedRpp}, in which the datapoints from Figure~\ref{fig:Rpp}
are divided by the curves from the same Figure, to form a new quantity, $R^\prime_{pp}$.
  Explicitly, the green circles on Figure~\ref{fig:EMCcorrectedRpp},
which compare multiplicity selections \#9 and \#4 are given by
\begin{align}
\label{eq:RppPrime}
&R^{\prime\left(\#9,\#4\right)}_{pp}\left(p_T\right)  \equiv \frac{\left.\frac{dn}{dp_T}\right|_{\#9}}{\left.\frac{dn}{dp_T}\right|_{\#4}}
  \times  \\
& \exp\left[\left(\frac{1}{2\left(N_{\#9}-1\right)}-\frac{1}{2\left(N_{\#4}-1\right)}\right)
     \left(\frac{2p_T^2}{\langle p_T^2\rangle}+\frac{\left(E-\langle E\rangle\right)^2}{\langle E^2\rangle-\langle E\rangle^2}\right)\right] , \nonumber
\end{align}
where the relevant quantities from Table~\ref{tab:pp} are used.  Again,
all particles are assumed to have pion mass.
Qualitative though it is, Figure~\ref{fig:EMCcorrectedRpp} raises the possibility that, when ``trivial'' EMCICs are accounted for,
the high-$p_T$ yield from high-multiplicity \pp collisions is {\it suppressed} relative to low-multiplicity collisions,
a trend in the same direction as that observed in \AuAu collisions.

In the hard sector, our estimates are mathematically and conceptually
 too simplistic to decide whether this implies ``jet quenching'' in high-multiplicity \pp collisions.
However, it is quite clear that conservation-induced phasespace restrictions might be sufficiently large in the hard sector, so that a
high-$p_T$ ``enhancement'' in high-multiplicity \pp collisions turns into a ``suppression,'' when these effects are accounted for.
Extracting physics messages (e.g. about mini-jet production or jet quenching) from the multiplicity evolution of \pp spectra
is a non-trivial task, in light of this potentially huge background effect.  At the very least, EMCICs should not be ignored, as they
usually are, when extracting physics messages.

\section{Summary and Discussion}
\label{sec:discussion}

The study of relativistic heavy ion collisions is, by its very nature, heavily dependent
on comparative systematics.  Physical models or hypotheses are most stringently tested when predictions
for a given observable are compared to measurements for a range of global collision conditions.
Even aside from specific models, much qualitative information may be gleaned simply through
study of the evolution of an observable as collision conditions-- quantified by global variables-- change.

Since the goal is to probe an interaction or transition characterized by a dimensionful scale (confinement length $\sim 1$~fm),
perhaps the most important global variable is event multiplicity, which on average reflects the size
of the system generated in the collision.  

Directly measurable is the multiplicity evolution of experimental observables.
This evolution is driven by (1) the evolution of the underlying physics-- which is of direct interest and (2)
kinematic phase-space restrictions (EMCICs)-- which are presumably less interesting.
It may be hazardous to ignore the latter effect and make inferences on the former, particularly since phase-space
restrictions have an obvious explicit multiplicity dependence.  
In this study, we have quantitatively estimated
the degree to which phase-space restrictions may affect physics inferences based on measured data.


We have focused on the multiplicity evolution of single particle spectra.  In previous 
published studies, analyses which have ignored EMCICs have inferred much from this evolution.  In particular,
there have been conclusions that spectra from central \AuAu collisions exhibit greater collective radial
flow than do those from peripheral \AuAu or \pp collisions.
Using an expression to approximately account for EMCIC effects, we have shown that the multiplicity evolution of
the spectra may be dominated by such effects, rather than any change in the underlying physics.

In particular, we have tested the extreme postulate that the driving physics, characterized by a parent distribution,
is {\it identical} for \pp collisions and \AuAu collisions of all centralities.
Since the parameters characterizing the parent distribution and the system multiplicity $N$ were fitted,
our test is not perfect.  Some multiplicity evolution of the parent distribution itself may exist, and may
not be easily separable from EMCICs.  Our point is that, with ``reasonable'' parameters, much of the data
systematics is readily understood in terms of a universal parent distribution in the soft sector, and
similar high-$p_T$ yield suppression in \pp and \AuAu collisions.

In the soft sector ($p_T\lesssim 1$~GeV/c) this postulate worked surprisingly well.
The changes in $m_T$ distributions, as the collision multiplicity is changed, are almost entirely due to EMCICs.
``Correcting'' the spectra for EMCICs, an approximate procedure along the lines of Fermi's Golden Rule,
reveals almost universal parent distributions.  

While the spectra themselves carry more information than fits
to the spectra, it was interesting to find that blast-wave fits to the ``EMCIC-corrected'' spectra show
that low multiplicity \AuAu collisions, and even \pp collisions, are characterized by very similar flow
and temperature values as for spectra from \AuAu collisions.  This contrasts strongly with previous conclusions
and assumptions about collectivity in small systems.
Blast-wave~\cite{Retiere:2003kf,Adams:2003xp} or modified Blast-wave~\cite{Tang:2008ud} fits which ignore
EMCICs, may yield unreliable results for low-multiplicity final states.

The same analysis of $p_T$ spectra of unidentified hadrons from multiplicity-selected \pp collisions yielded similar results,
though the multiplicity evolution of the spectra was only roughly explained by our postulate.  This is to
be expected, for several reasons.  Firstly, our approximate expression to account for EMCICs was based on the
central limit theorem, which begins to break down for the very small multiplicities involved.  Secondly, the
lack of particle identification led to a simple assumption that all particles were pions.  Nevertheless,
it was clear that EMCICs can go a long way towards explaining the multiplicity evolution of the $p_T$ spectra
in the soft sector.

EMCIC effects on momentum distributions are expected to be large at higher $p_T$, where a single particle may consume
much of the total available energy.
However, the approximations behind our EMCIC factor should begin to break down at high $p_T$.  Unlike our results in
the soft sector, we would be on shaky ground to draw firm conclusions from our studies in the hard sector.
Nevertheless, we applied our formalism to obtain a rough estimate the magnitude of restricted phase-space effects at high $p_T$.

Firstly, we immediately realized that the well-known ``high-$p_T$ suppression'' for central \AuAu collisions can not be explained
by EMCICs, as these effects would cause the opposite behavior (i.e. ``high-$p_T$ enhancement'') from what is experimentally observed.
Thus, our postulate fully breaks down at high $p_T$-- there is a difference in the {\it physics} (parent distribution) in the hard sector.

Turning to the multiplicity-evolution of $p_T$ spectra from \pp collisions, however, the measured effect goes in the same direction
as that expected from EMCIC effects.
Still keeping in mind the caveats behind our expression at high momentum, we estimated that the high-$p_T$
enhancement expected from EMCICs should be at least as large as that observed in the data.  Again, we do not conclude, but
{\it suggest} that the multiplicity-evolution of the parent distributions in \pp collisions might in fact reveal a high-$p_T$ suppression
for high multiplicity collisions, reminiscent of the effect measured in heavy ion collisions.

\section{Conclusions and outlook}
\label{sec:outlook}

Our results suggest that
the multiplicity evolution of the soft portion of the $p_T$ spectra in collisions at RHIC is
dominated by phase-space restrictions.  Effects due to actual changes in physics (the parent distribution)
are subdominant.  This suggests one of two possibilities.

Firstly, one may take the common assumption that the physics underlying the soft particles from
$A+A$ and \pp collisions is quite different, say
bulk behavior versus string breaking, respectively.  In this case, our results suggest that
single-particle spectra are too insensitive to distinguish very different physics scenarios,
and physics conclusions (say, radial flow in $A+A$ collisions) based on them are questionable.

On the other hand, the single-particle spectra may well reflect the underlying physics.
If energy and momentum conservation effects are taken into account, the low-$p_T$ spectra
indicate that \pp collisions display as much collective radial flow as do \AuAu collisions.  In the larger
system, this collective behavior is usually considered to arise from a (perhaps only partially) thermalized
{\it bulk} system.

The question naturally arises: isn't it impossible for a system as small as that created in a \pp
collision to form even a partially thermalized bulk system which develops flow?  The answer is not obvious.  
After all, estimates set the timescale for {\it complete}
thermalization in central \AuAu collisions below 1~fm/c~\cite{Kolb:2003dz,Huovinen:2006jp}, via
a mechanism that may be driven more by fluctuating color fields than by classical rescattering 
processes~\citep[][,and references therein]{Mrowczynski:2007hb}.
Perhaps the possibility that similar processes have sufficient time to thermalize a system on the scale of
$\sim 1$~fm should not be dismissed out of hand.

Indeed, in the literature one finds frequent suggestions~\cite{VanHove:1982vk,Breakstone:1986xs,Levai:1991be,Alexopoulos:2002eh,Alexopoulos:1990hn,Steinberg:2004wx},
based on single-particle spectra, that high energy particle collisions generate flowing bulk systems
and perhaps even Quark-Gluon Plasma; see also the recent review by Weiner~\cite{Weiner:2005gp}.  
By partially removing the obscuring effects of EMCICs, we have more directly compared proton collisions to
heavy ion collisions (at the same energy and measured with the same detector), for which a flow-based
interpretation is generally well accepted.

If a bulk system {\it is} created in \pp collisions, might it ``quench'' jets as the medium does in \AuAu collisions?
This was, after all, the original proposition of Bjorken~\cite{Bjorken:1982tu}.  The signature of such quenching would be a suppression
of particle yields at high $p_T$ in high-multiplicity collisions, relative to those at lower multiplicity.
While our formalism is insufficiently reliable at high $p_T$ to draw firm quantitative conclusions, such a suppression
may possibly be present, though obscured by EMCICs in measured spectra.

Increased focus on the relationship between large and small systems created in ultrarelativistic collisions
is called for.  Experimental programs at the Large Hadron Collider will very soon open up important avenues in this
study.  In particular, the experiments will measure first \pp collisions at record collision energies, with event
multiplicities similar to $Cu+Cu$ or semi-peripheral \AuAu collisions at RHIC.  Soft sector $p_T$ distributions will likely
be among the first observations reported.  Later, with identical acceptance and techniques,
the same experiments will then measure much larger systems created in $Pb+Pb$ collisions.  The direct comparison afforded
by this data should help answer the question of whether a bulk system created in hadronic collisions is qualitatively
different than that created in collisions between the heaviest ions, or merely a smaller version of it.

The nature of relativistic heavy ion studies depends upon comparison of ``small'' and ``large'' collision systems,
each of which may be driven by distinct, non-trivial physics processes.  In performing such comparisons, we must not
neglect the ``trivial'' effect of energy and momentum conservation, and its explicit dependence on collision size.

\begin{acknowledgments}
We wish to thank Drs.~Pawe\l ~Danielewicz, 
Sean Gavin,
Ulrich Heinz,
Peter Jacobs,
Declan Keane,
Scott Pratt,
Sergei Voloshin, 
 and 
Bill Zajc,
for important suggestions and insightful discussions,
and Mr. Evan Askanazi for some numerical calculations.

This work supported by the U.S. National Science Foundation under Grant No. PHY-0653432.
\end{acknowledgments}

\appendix

\section{EMCIC factors for rapidity- and angle-integrated $p_T$ distributions}
\label{sec:rapidityAppendix}

Equation~\ref{eq:EMCIC2} gives the EMCIC correction factor to the triple differential spectrum $\tilde{f}\left(p\right)$.
Experimental measurements often report $p_T$ distributions integrated over angle and a range of rapidity, i.e.
\begin{align}
\label{eq:fcpt}
&\tilde{f}_c\left(p_T\right) \equiv \frac{1}{4\pi\cdot y_{max}} \int_0^{2\pi}d\phi\int_{-y_{max}}^{y_{max}} dy \tilde{f}_c\left(p_x, p_y, p_z, E\right) . 
\end{align}

\begin{figure}[t!]
{\centerline{\includegraphics[width=0.5\textwidth]{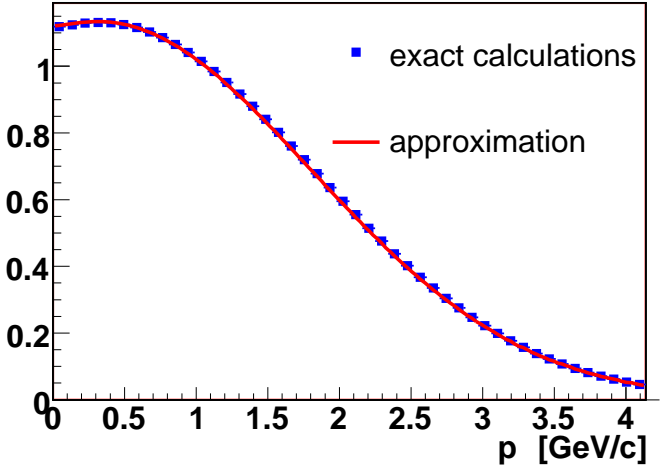}}}
\caption{(Color online)  EMCIC factor calculated using the numerical averaging of Equation~\ref{eq:rap1} and the
approximation of Equation~\ref{eq:rap2}.
\label{fig:verify}}
\end{figure}

In the absence of a triple-differential measurement, we consider azimuthally-symmetric distributions, and
$\langle p_x^2\rangle=\langle p_y^2\rangle=\langle p_T^2\rangle/2$.  At midrapidity at RHIC, it is
reasonable also to assume a boost-invariant parent distribution.  In this case, only part of the
EMCIC factor remains in the rapidity integral:
\begin{align}
\label{eq:rap1}
\tilde{f}_{c}\left(p_T\right) &= \tilde{f}\left(p_T\right) \cdot
    \left(\frac{N}{N-1}\right)^{2} \exp\left[\frac{-p_T^2}{\left(N-1\right)\langle p_T^2\rangle}\right]
\times \nonumber \\
& \frac{1}{2y_{max}} \int_{-y_{max}}^{y_{max}} dy
 \exp\left[
  \frac{-1}{2\left(N-1\right)}\left(
\frac{p_z^2}{\langle p_z^2\rangle} + \right.\right.  \\ 
&\left.\left.  \frac{E^2}{\langle E^2\rangle-\langle E\rangle^2}
- \frac{2E\langle E\rangle}{\langle E^2\rangle-\langle E\rangle^2}
+ \frac{\langle E\rangle^2}{\langle E^2\rangle-\langle E\rangle^2}
\right)
\right] . \nonumber
\end{align}

To arrive at a closed form for our EMCIC factor, we approximate the average of the
exponential with the exponential of the average, i.e.
\begin{align}
\label{eq:rap2}
\tilde{f}_{c}\left(p_T\right) &= \tilde{f}\left(p_T\right) \cdot
    \left(\frac{N}{N-1}\right)^{2} \times  \nonumber \\
&  \exp\left[
- \frac{1}{2\left(N-1\right)}
\left(\frac{2p_T^2}{\langle p_T^2\rangle}+\frac{\overline{p_z^2}}{\langle p_z^2\rangle} \right.\right.  \\ 
&\left.\left. + \frac{\overline{E^2}}{\langle E^2\rangle-\langle E\rangle^2}
- \frac{2\overline{E}\langle E\rangle}{\langle E^2\rangle-\langle E\rangle^2}
+ \frac{\langle E\rangle^2}{\langle E^2\rangle-\langle E\rangle^2}
\right)
\right] . \nonumber
\end{align}
This expression is reproduced in Equation~\ref{eq:EMCfpt}.

Here, the rapidity-averaged quantities are
\begin{align}
\label{eq:pz2avg}
\overline{p^{2}_{z}} \equiv \frac{1}{2y_{max}}\int^{y_{max}}_{-y_{max}} p_z^2 dy
     =  m^{2}_{T} \left(\frac{\sinh(2 y_{max})}{4 y_{max}} - \frac{1}{2}\right) 
\end{align}
\begin{align}
\label{eq:en2avg}
\overline{E^{2}}  \equiv \frac{1}{2y_{max}}\int^{y_{max}}_{-y_{max}} E^2 dy
      =  m^{2}_{T} \left(\frac{\sinh(2 y_{max})}{4 y_{max}} + \frac{1}{2}\right) 
\end{align}
\begin{align}
\label{eq:enavg}
\overline{E} \equiv \frac{1}{2y_{max}}\int^{y_{max}}_{-y_{max}} E dy
      = m_{T} \frac{\sinh(y_{max})}{y_{max}} .
\end{align}

The approximation used in going from Equation~\ref{eq:rap1} to~\ref{eq:rap2} is well-justified
for typical numerical values used in this study.  Figure~\ref{fig:verify} shows a numerical
integration of the EMCIC factor from Equation~\ref{eq:rap1} (labeled ``exact'') and Equation~\ref{eq:rap2} (``approximation'')
for values indicated in the Figure.

\section{Region of applicability for the EMCIC formula}
\label{sec:ApplicabilityAppendix}

The exact expression for the phase space integral of Eq.~\ref{eq:EMCIC1} was approximated by that in
Eq.~\ref{eq:EMCIC2} through an appeal to the Central Limit Theorem.  
Discrepancies between the exact expression and the approximate Gaussian functional form
will become more apparent in the tails of the distribution.  
For example, our approximate phase space suppression function never vanishes, thus
permitting a tiny but finite probability for a particle to carry more energy than that
of the entire system!
In this Appendix, we perform simple numerical calculations with the
\genbod computer program~\cite{James:1968gu}, to estimate 
the range of quantitative reliability of Equation~\ref{eq:EMCIC2}.

\begin{figure}[t!]
{\centerline{\includegraphics[width=0.5\textwidth]{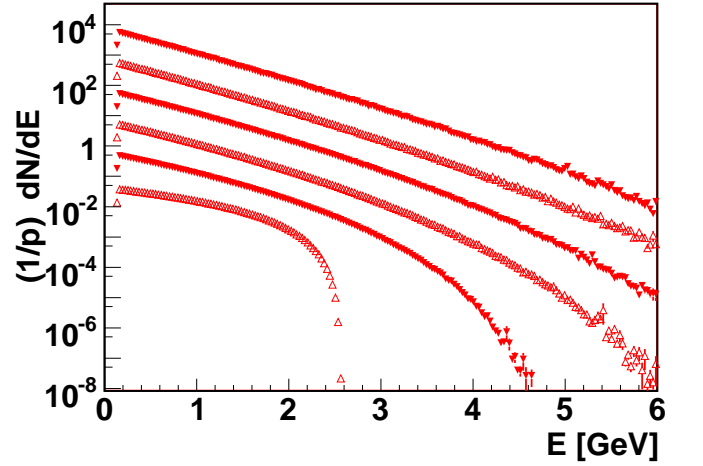}}}
\caption{(Color online) $\frac{1}{p}\frac{dN}{dE}$ obtained from  \genbod events run for 
the same average energy ($\langle E\rangle_c=1~GeV$) but different multiplicities: $N=5$, 10, 15, 20, 30, 40 pions.
\label{fig:GenBodSpectraNdep}}
\end{figure}

Given a total energy $E_{\rm tot}$, multiplicity $N$ and list of particle masses, \genbod produces phasespace-weighted events of $N$ 4-momenta by filling Lorentz-invariant phase space according to the Fermi distribution,
\begin{align}
\label{eq:fdist}
&\tilde{f} \equiv 2E \frac{d^3N}{dp^3} = \frac{1}{2\pi p}\frac{dN}{dE}\propto e^{-E/\zeta}.
\end{align}
where $\zeta$ characterizes the slope of the energy distributions.  
Since it is $(1/p)\cdot dN/dE$ which is exponential and not $(1/p)\cdot dN/dE$, the inverse slope $\zeta$ should not be considered a ``temperature,'' but only a parameter characterizing the parent distribution.

As a result, generated particles in an event are correlated only by energy and momentum conservation.
Thus, EMCIC effects on the calculated single-particle spectrum, $\tilde{f}_c\left( p\right)$,
are given precisely according to Equation~\ref{eq:EMCIC1}.

To evaluate the region of validity of Equation~\ref{eq:EMCIC2}, we use Eq.~\ref{eq:fdist} as a parent distribution, $\tilde{f}\left( p\right)$. Results
of this exercise are presented on Figure~\ref{fig:GenBodSpectraNdep} which shows energy spectra from \genbod events with the same average energy per particle
$\langle E\rangle_c=E_{\rm tot}/N=1$~GeV, but different multiplicity $N$.
As expected, in the limit of large $N$, $\tilde{f}_c\left(p\right) \rightarrow \tilde{f}\left(p\right)$,
and it is clear that the plotted distribution is increasingly well-described by an
exponential, as $N$ increases.

\begin{figure}[t!]
{\centerline{\includegraphics[width=0.5\textwidth]{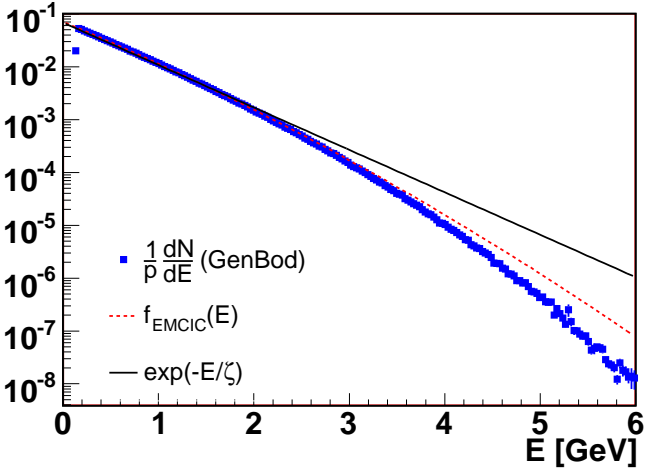}}}
\caption{(Color online) 
Blue points are $\frac{1}{p}\frac{dN}{dE}$ obtained from  \genbod events run for $N=20$, $\langle E\rangle=1$~GeV.
Black solid curve is an exponential, the assumed parent distribution; c.f. Equation~\ref{eq:fdist}.  Red dashed
curve is the exponential times the EMCIC factor, as per Equation~\ref{eq:fcdistfit}.
\label{fig:GenBodFs}}
\end{figure}

\begin{figure}[t!]
{\centerline{\includegraphics[width=0.5\textwidth]{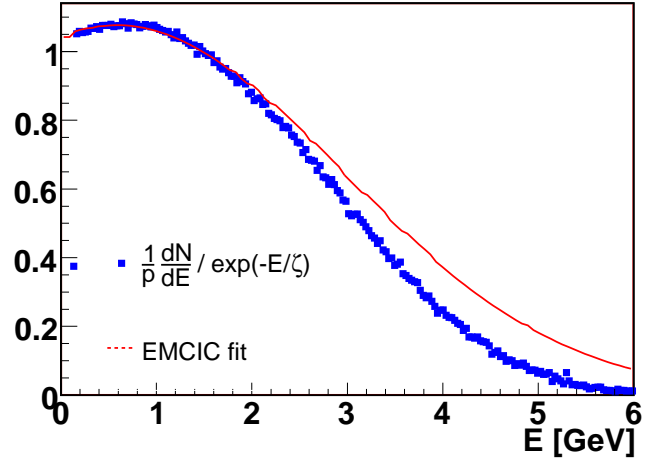}}}
\caption{(Color online)  
Blue points are $\frac{1}{p}\frac{dN}{dE}$ obtained from  \genbod events run for $N=20$,$\langle E\rangle=1$~GeV,
divided by $\exp\left(-E/\zeta\right)$; i.e. the blue points from Fig.~\ref{fig:GenBodFs} divided by the black
full curve from the same figure.
Red dotted line is the EMCIC factor; i.e. 
the red dotted curve from Fig.~\ref{fig:GenBodFs} divided by the black
full curve from the same figure.
\label{fig:GenBodFit}}
\end{figure}

It is appropriate here to point out why we wish to identify the parent distribution in the first place,
rather than following the procedure outlined in Section~\ref{sec:postulate}.  There, the
parent distribution cancels when taking the ratio of two measured spectra $\tilde{f}_{c,1}/\tilde{f}_{c,2}$,
using the postulate that the parent distributions $\tilde{f}_{1}$ and $\tilde{f}_{2}$ are identical.
In contrast, the parent distributions for the different \genbod spectra shown in Figure~\ref{fig:GenBodSpectraNdep} are
assuredly {\it not} the same.  Those spectra came from event samples having the same $\langle E\rangle_c$ 
(c.f. Eq.~\ref{eq:MeasuredAverages}), and thus {\it different} $\langle E\rangle$ (c.f. Eq.~\ref{eq:averages}),
implying different parents.

Having at hand a functional form for the \genbod parent distribution, we may test our approximate
formula for the phasespace modification factor, by fitting the calculated spectrum according to
\begin{align}
\label{eq:fcdistfit}
\frac{dN_c}{dE} &= A \cdot p \cdot e^{-E/\zeta} \quad \times \\ 
& \left( \frac{N}{N-1}  \right)^2
  \exp\left[
\left(-\frac{1}{2\left(N-1\right)}\right)
\left(\frac{3p^2}{\langle p^2\rangle}+ \right.\right. \nonumber \\
&\left.\left. + \frac{E^2}{\langle E^2\rangle-\langle E\rangle^2}
- \frac{2E\langle E\rangle}{\langle E^2\rangle-\langle E\rangle^2}
+ \frac{\langle E\rangle^2}{\langle E^2\rangle-\langle E\rangle^2}
\right)
\right], \nonumber
\end{align}
where we used the fact that \genbod generates particles isotropically so that $<p_x^2>=<p_y^2>=<p_z^2>=\frac{1}{3}<p^2>$.
Since $N$ is a known quantity, and $\langle E\rangle$, $\langle E^2\rangle$ and $\langle p^2\rangle$ may be directly calculated from $\zeta$, the fit
of Equation~\ref{eq:fcdistfit} has only two parameters: the overall normalization $A$, which is unimportant to us, and $\zeta$, which characterizes the
parent distribution.

The results are shown in Figure~\ref{fig:GenBodFs} and, for better 
detail, in Figure~\ref{fig:GenBodFit}.  For the case here, which
is typical of that in the data, we see that our approximation begins 
to break down for particle energies $E \gtrsim 2\div 3 \langle E \rangle$.  Above this
range, our approximation (e.g. Equation~\ref{eq:ratio}) should only be taken qualitatively.

\end{document}